\documentclass[runningheads]{llncs}
\usepackage{graphicx}
\usepackage{booktabs}
\usepackage{multirow}

\begin{document}
\title{Archify: A Recommender System of Architectural Design Decisions}
\titlerunning{Archify: A Recommender System of Architectural Design Decisions}
%

\author{Breno Cruvinel Marinho\inst{1}\orcidID{0000-0002-7584-2646} \and
Renato Bulcão-Neto\inst{1}\orcidID{0000-0001-8604-0019} \and\\
Valdemar Vicente Graciano Neto\inst{1}\orcidID{0000-0003-2190-5477}}
\authorrunning{Marinho et al.}
%
\institute{Instituto de Informática -- Universidade Federal de Goiás\\ Goiânia, Brazil}
\maketitle              
\begin{abstract}
Software architectures play a critical role in software quality assurance. However, small and medium companies (SMC) often suffer from the absence of professionals with skills and expertise in software architecture. That situation potentially affects the final quality of the software products and pressures projects budget with extra costs with consulting. This paper presents a recommender system of architectural design decisions called Archify. The goal is to support SMC companies in part of the effort of architecturally designing their products. Archify implements a wizard-styled interface that guides the developer or project manager through a set of specific questions. While the user answers these questions, Archify buffers a set of corresponding architectural decision recommendations. As the final result, the system recommends a set of architectural decisions matching the project's needs according to the requirements (as provided by the user) of the software under development. Nineteen professionals from academia and industry evaluated Archify through two surveys. The findings reveal that 94.7\% of the participants approved Archify as a supporting tool. Respondents also highlighted the lack of tools supporting software architecture design, remarking the relevance of the proposed system.


\end{abstract}
\section{Introduction}

Software Architecture (SA\footnote{For simplicity, the acronym SA will be interchangeably used to express both singular and plural forms: software architecture and software architectures.}) is the backbone of any well-succeeded software system and plays an important role in software quality assurance \cite{michael2009verification}. According to survey results with IT practitioners, among 126 participants, 27\% worked as software architects throughout small, medium and large companies from 39 different countries \cite{survey:2018}. From another study with 395 developers, authors extracted perceptions and state that ``even the software architect's role is often not clearly defined'' \cite{melo:2016}. As a matter of fact, small and medium-sized companies (SMC) often do not have a professional playing a software architect's role. Consequently, the software products delivered by those companies can suffer from a decreased quality \cite{EstudoQualitativo}. Therefore, in companies where there are no software architects, a tool that gathers expert knowledge on the SA design process could help to overcome that problem.


In that context, the main contribution of this paper is twofold: the Recommender System (RS) for SA called Archify and the results of two surveys conducted. Archify advances state of the art and practice while supporting both the most and the least experienced professionals by suggesting architectural design alternatives for a given software development project. From an intuitive wizard-based process, the user is presented with twelve questions about specific concerns to extract the characteristics of the project being developed. In the end, according to the answers collected, the Archify system processes and recommends a set of architectural alternatives divided into three categories, which are: (i) architectural styles, (ii) architectural tactics, and (iii) technologies. The surveys evaluated it according to (i) the set of questions proposed to guide the recommendation process and (ii) the results delivered by the proposed tool. Results revealed that the system's mean evaluation was 4 out of 5 and the lack of supporting tools for the software design phase, which corroborates this research and corresponding tool's importance.

This paper is structured as follows: Section 2 presents the background and related work. Section 3 presents the Archify system. Section 4 presents the evaluation and, Section 5 discusses the results. Finally, Section 6 presents the final remarks and future work.

\section{Background}

An SA can be considered the result of a set of design decisions \cite{jansen2005software}, i.e, the result of a decision-making process 
that will shape the final product. Architectural design decisions are mainly concerned with the (i) prioritized quality attributes (QA), (ii) architectural styles/patterns used to satisfy the QA, (iii) application domain of the system and (iv) COTS components and technologies 
\cite{jansen2005software}. Architectural tactics, which are strategies to enhance QA (e.g. resources redundancy to enhance availability), are also regarded as architectural decisions \cite{Bass:2012}.

The set of architectural decisions that compose a SA is the result of the SA design process. The classical process presented by Hofmeister et al. consists of three main activities \cite{hofmeister2007general}: (i) architectural analysis, (ii) architectural synthesis, and (iii) architectural evaluation. The important concerns, i.e., the architecturally significant requirements (ASRs), are identified during the architectural analysis. Potential solutions (architectural alternatives based on architectural decisions established in the first step) are developed during architectural synthesis. Architectural evaluation is then a validation step, where the identified solutions are checked against the identified ASRs. The result of the entire process is an agreed architecture, which is used as input for the next steps in the software development life cycle \cite{weinreich2012towards}.
 
Obtaining a set of architectural decisions is a difficult endeavor \cite{paris2013difficulty} and the conduction of a software architectural design process is expensive, complex and often requires the presence of a software architect. Hence, automating parts of the process can save time and budget. Recommender systems (RS) with expert knowledge embedded could then be used to automate or semi-automate part of those activities. 
Even in the context of SA, recommender systems have emerged for dealing, for instance, with architectural violations \cite{terra:2015,terra2}. These types of systems can be classified as follows \cite{RecommenderSystems}: (i) \textbf{Collaborative filtering}, in which recommendations are based on the relationships between users whose preferences can be similar to produce predictions for other users; (ii) \textbf{Content-based}, where recommendations are based on the historical data of the \textit{liked} items. Each liked item has some keywords assigned to it and allows the system to suggest similar items; (iii) \textbf{Knowledge-based}, an approach focused on recommending content based on user inputs. The system recommends results related to the queries made by the user. This RS can be of two types: \textit{constraint-based} and \textit{case-based}. In  the former, the system requires the user for input each time and combines it with internal rules for presenting recommendations, while in the latter, previous queries are used as input information to be combined with the next queries, (iv) \textbf{Demographic}, in which user demographic data are used to identify users behaviors. Demographic groups are identified, mapped and analysed for recommending content. The users will receive content recommendations based on content positively evaluated by other users with similar demographic information; and (v) \textbf{Hybrid}, which explores different application approaches to work together to minimize weaknesses and enhance recommendations. RS that use this approach tend to be more robust and complex. 
\\\\
\noindent \textbf{\underline{Related Work.}} RS have already been proposed for SA domain \cite{Identifying,Brandner,ArchReco}. Most of them are Knowledge-based RS, i.e., the recommendation is underpinned by a process where the system collects some information from the user and recommends suggestions based on that interaction. 


Anish et al. \cite{Identifying}  
created the tool called ArcheR to deal with functional requirements and their impact on SA design. ArcheR (i) automates the identification of architecturally significant functional requirements, (ii) classifies them into categories based on the different types of architectural impact they may have, (iii) recommends probing questions that the analyst should consider to produce a more complete analysis of the requirements specification and (iv) recommends possible architectural solutions in response to the architectural impact. However, they do not recommend tactics to reinforce QA, nor use a question-based style or generates documentation.

 Brandner and Weinreich proposed a RS that, based on an Architecture Knowledge Base (AKB), delivers recommendations for technologies, IDE and architectural styles  \cite{Brandner}. The authors follow five strategies for recommendations: similar software, similar context, similar design space, related design space and design process. The user chooses these strategies and searches for architectural models through search tags (keywords). The final result presented by the system is a list of architectural decisions related to the informed tags. 


Sielis et al. proposed the tool ArchReco, an SA design tool enhanced by context aware recommendations for design patterns \cite{ArchReco}. The tool was developed based on a study that identified the needs of SA professionals in terms of supporting mechanisms to assist in the process of designing SA. 
ArchReco is a desktop tool developed in Java 
that allows users to: (i) define the SA using diagram while offering suggestions for design patterns and (ii) collaborate with each other during this process. In the time of the publication, the tool was still under development. 


Other studies were also found, such as \textit{ArchE} \cite{Towards} and Gastón Márquez \cite{MarquesGaston}. However, ArchE is not open to the public, what prevented an effective testing it and Gastón Márquez did not provide any tools. The studies proposed by Brandner and Weinreich \cite{Brandner} and the tool ArcheR \cite {Identifying} are the main related works and have similarities in their recommendations. However, Archify advances the state of the art by gathering all the main functional characteristics an RS for SA should have in one single tool. In the next section, we present Archify RS.

\section{The Archify Recommender System}

    Archify is a web knowledge-based recommender system and relies on the perspective of an SA as a set of architectural decisions based on architectural styles, architectural tactics and technologies. A video presenting the tool is available externally\footnote{https://vimeo.com/447360530} as well as the tool itself\footnote{https://archify.com.br/} for navigation and use. Archify assists users on part of the architectural analysis and architectural synthesis steps prescribed by Hofmeister \cite{hofmeister2007general}.
    The architectural model resulting from the decision-making process is one among a wide variety of possible combinations of models and decisions. Archify performs several successive pruning in the different possibilities for each of the choices made in the process, combining the possibilities suggested in each dimension of an architecture and delivering a set of quality attributes, corresponding architectural styles, architectural tactics and technologies.

    The Archify \textit{back-end} was developed in \textit{PHP \footnote{https://www.php.net/}}, running on a \textit {Apache
    }. \textit{MariaDB
    } database, a \textit{mySQL
    } distribution, was used for data storage, whilst  the \textit{front-end} was created on \textit{HTML5
    }. Archify is based on \textit{MVC
    } to maintain the decoupling between the parts. Archify runs on a \textit {wizard
    } model, which acts as a guide that conducts the user through a series of closed-ended questions with pre-defined architectural decisions associated to each one of the answers. 
    The questions used during the recommendation process are stored in a database. Hence, the architecture is highly maintainable and cohesive, and the user interaction works independently of the backend. The addition of new concerns and respective questions consists of just adding them to the database.


    The SA design process supported by Archify starts with the architectural analysis. 12 questions about recurrently important concerns in software development projects drive the process. Table \ref{tab:perguntas} presents the recommendation process questions (RPQ). They were elaborated to identify several characteristics of the software, allowing the suggestion of corresponding architectural alternatives. The questions concern the software domain (question 1) and some quality attributes and non-functional requirements \footnote {In this paper, we assume that some non-functional requirements are not quality attributes, such as technology restrictions and legislation that restricts the operation of certain software systems.} such as distribution (question 2), scalability (question 3) and technologies (questions 4 and 5), interoperability (question 6) and other aspects, which can be seen in Table \ref{tab:perguntas}.

\begin{figure*}[!ht]
\centering
\includegraphics[scale=0.30]{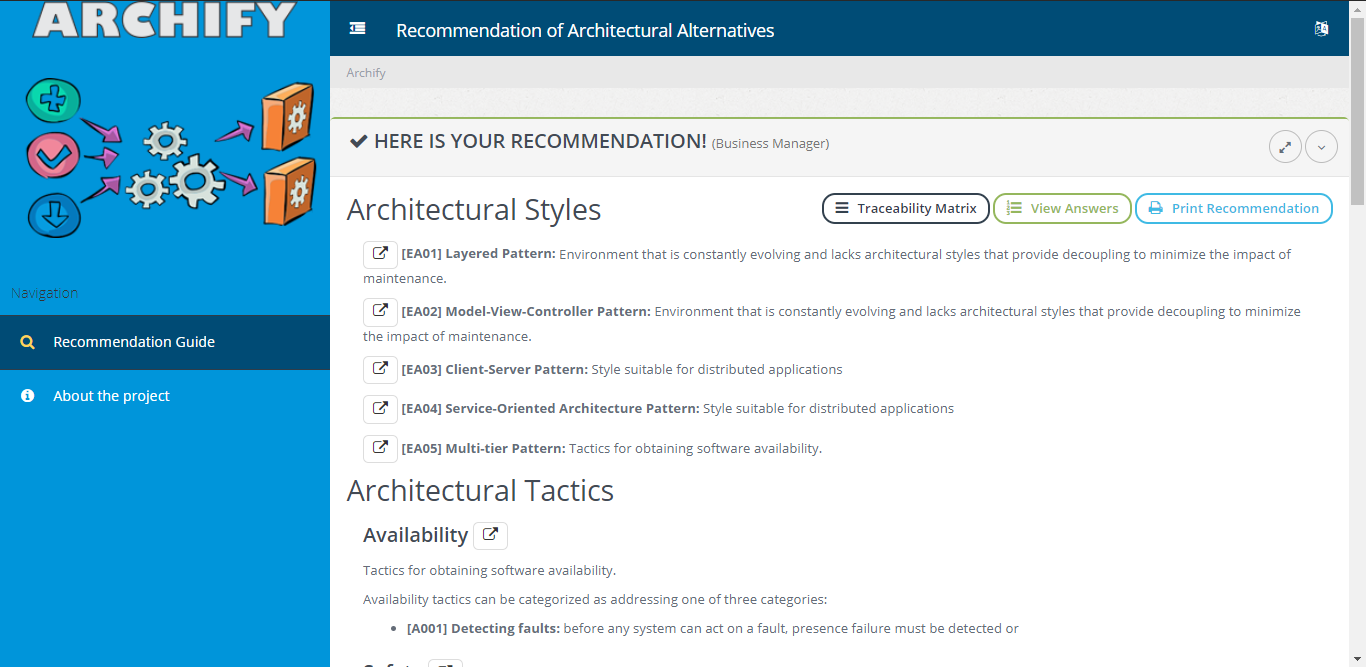}
\caption{Archify recomendation result screen.}
\label{fig:printscreen-archify}
\end{figure*}

    Finally, after the questionnaire ends, the recommendations for architectural alternatives are exhibited. The result contains the list of recommendations classified into three categories: (i) architectural styles, (ii) architectural decisions and (iii) technologies. In addition, an explanatory text has been added for each recommended item. A demonstrative figure is presented for each suggested architectural style to maximize the understanding. The recommendations can then be exported in PDF format with the purpose of being an artifact that can be used as part of the SA documentation. Figure \ref{fig:printscreen-archify} depicts architectural alternatives recommended by Archify in response to the user's input information.

\begin{table}[!ht]
\caption{Archify elaborated questions}
\label{tab:perguntas}
\centering
\begin{tabular}{|l|}
\hline
RPQ1. What is the software domain? \\ \hline
\begin{tabular}[c]{@{}l@{}}RPQ2. Does this software have the characteristics of a distributed application?\end{tabular} \\ \hline
\begin{tabular}[c]{@{}l@{}}RPQ3. The number of users that the software must serve, is a known number or \\does the system provide resilience and scalability, that is, capacity to modify the\\ amount of resources provided from varying demand?\end{tabular} \\ \hline
\begin{tabular}[c]{@{}l@{}}RPQ4. Does the team that will develop the software already have expertise in any\\ technology?\end{tabular} \\ \hline
\begin{tabular}[c]{@{}l@{}}RPQ5. Does The team that will develop the software has expertise in what type\\ of database?\end{tabular} \\ \hline
\begin{tabular}[c]{@{}l@{}}RPQ6. Should the Software perform interaction (s) with other software (s)?\end{tabular} \\ \hline
\begin{tabular}[c]{@{}l@{}}RPQ7. Regarding the data that will be transmitted to another software: do the\\ data types follow strict typing and validation rules?\end{tabular} \\ \hline
\begin{tabular}[c]{@{}l@{}}RPQ8. Regarding availability, if there is temporary unavailability of the software,\\ can users be at risk, be hurt or have financial or other losses?\end{tabular} \\ \hline
\begin{tabular}[c]{@{}l@{}}RPQ9. Regarding the software maintainability, are there prospects for frequent\\ changes/evolutions in the system?\end{tabular} \\ \hline
\begin{tabular}[c]{@{}l@{}}RPQ10.Regarding the security, will the software store important data of\\ interest to third parties?\end{tabular} \\ \hline
\begin{tabular}[c]{@{}l@{}}RPQ11. Regarding the usability, does the software need user efficiency with\\ respect to self-learning, minimizing the impact of errors or related?\end{tabular} \\ \hline
\begin{tabular}[c]{@{}l@{}}RPQ12. Is the elasticity of the database, i.e., the ability of software to scale its \\storage technology as well as stored data, an important factor?\end{tabular} \\ \hline
\end{tabular}
\end{table}
    
    Archify also provides two other resources: (i) a traceability matrix (that shows how the provided decisions are mapped onto the answers delivered in each question) and (ii) a graphical option that links the answer taken by the user with the recommendations provided by the tool. 
    Thus, the tool also has an educational bias that allows the user to learn how the suggested architectural alternatives were obtained. 

Archify has a knowledge base with possible architectural alternatives. Recommendations are linked to each answer option for the questions asked. In the Archify pilot version presented in this paper, the recommendations were based on the literature \cite{Bass:2012}. Table \ref{tab:recomendacoes} exhibit the possible recommendations for each architectural alternative of each possible answer.

\begin{table}[!ht]
\caption{Recommendations for each question and answer.}
\scriptsize
\label{tab:recomendacoes}
\centering
\begin{tabular}{|l|l|l|}
\hline
\multicolumn{1}{|c|}{RPQ1} &
  Business &
  \begin{tabular}[c]{@{}l@{}}Layered Pattern\\ Model-View-Controller Pattern\end{tabular} \\ \cline{2-3} 
\multicolumn{1}{|c|}{} &
  Academic &
  Layered Pattern \\ \cline{2-3} 
\multicolumn{1}{|c|}{} &
  Hospital &
  \begin{tabular}[c]{@{}l@{}}Layered Pattern\\ Service-Oriented Pattern\end{tabular} \\ \cline{2-3} 
\multicolumn{1}{|c|}{} &
  \begin{tabular}[c]{@{}l@{}}Real-time game\end{tabular} &
  Real-Time Agent \\ \cline{2-3} 
\multicolumn{1}{|c|}{} &
  \begin{tabular}[c]{@{}l@{}}Web Conference / Stream (audio/video)\end{tabular} &
  Peer-to-Peer Pattern \\ \cline{2-3} 
\multicolumn{1}{|c|}{} &
  Other &
  - \\ \hline
{RPQ2} &
  Yes &
  \begin{tabular}[c]{@{}l@{}}Client-Server Pattern\\ Service-Oriented Pattern\end{tabular} \\ \cline{2-3} 
 &
  No / Don't know &
  - \\  \hline
RPQ3 &
  Yes &
  - \\ \hline
 &
  No &
  \begin{tabular}[c]{@{}l@{}}Publish-Subscribe Pattern\\ Clusters\end{tabular} \\ \hline
 &
  Don't know &
  - \\ \hline
{RPQ4} &
  PHP &
  PHP Framework - Laravel \\ \cline{2-3} 
 &
  C &
  C Framework - ASP.NET MVC \\ \cline{2-3} 
 &
  Java &
  Java Framework - Spring MVC \\ \cline{2-3} 
 &
  Python &
  Python Framework - Django \\ \cline{2-3} 
 &
  React Native &
  React-Native \\ \cline{2-3} 
 &
  \begin{tabular}[c]{@{}l@{}}None of these\end{tabular} &
  - \\ \hline
RPQ5 &
  SQL &
  SQL \\ \hline
 &
  NoSQL &
  NoSQL \\ \hline
 &
  Both &
  \begin{tabular}[c]{@{}l@{}}SQL\\ NoSQL\end{tabular} \\ \hline
\multirow{3}{*}{RPQ6} &
  Yes &
  Service-Oriented Pattern \\ \cline{2-3} 
 &
  No / Don't know &
  - \\ \hline
\multirow{3}{*}{RPQ7} &
  Yes &
  SOAP \\ \cline{2-3} 
 &
  No &
  REST \\ \cline{2-3} 
 &
  Don't know &
  - \\ \hline
\multirow{3}{*}{RPQ8} &
  Yes &
  \begin{tabular}[c]{@{}l@{}}Multi-tier Pattern\\ Clusters\\ Availability\end{tabular} \\ \cline{2-3} 
 &
  No / Don't know &
  - \\ \hline
\multirow{3}{*}{RPQ9} &
  Yes &
  \begin{tabular}[c]{@{}l@{}}Layered Pattern\\ Model-View-Controller Pattern\end{tabular} \\ \cline{2-3} 
 &
  No / Don't know &
  - \\ \hline
\multicolumn{1}{|c|}{\multirow{3}{*}{RPQ10}} & Yes     & Safety   \\ \cline{2-3} 
\multicolumn{1}{|c|}{}                     & No / Don't know    & -           \\ \hline
\multirow{3}{*}{RPQ11}                       & Yes     & Usability \\ \cline{2-3} 
                                           & No / Don't know     & -           \\ \hline
\multirow{3}{*}{RPQ12}                       & Yes     & NoSQL       \\ \cline{2-3} 
                                           & No     & SQL         \\ \cline{2-3} 
                                           & Don't know & -           \\ \hline  
\end{tabular}
\end{table}

The design of a SA is complex, which makes unfeasible to map the questions in the form of a decision tree, which could lead to a combinatorial explosion of possibilities. Recommendations are made by processing the response buffer, i.e., the answers are buffered to provide a final result and the order of the questions is not important. Therefore, it is necessary to define some business rules to recommend architectural alternatives according to the software requirements. 

        A priority index was assigned for each question. Thus, a system script takes into account the priority of the question to define which recommendation should be made. This is necessary because there are questions (such as RPQ5 and RPQ12) that can be classified as contradictory. For instance: if the developer team does not have expertise with NoSQL technology (RPQ5) and the project needs data elasticity (RPQ12), the use of NoSQL will not be recommended, since question 12 is prioritized over question 5 (a business rule defined for Archify). The next section presents results of the evaluation of Archify.

\section{Evaluation}

This section brings details on the evaluation of Archify tool. The methodology used to develop Archify includes six steps, as follows: (i) literature review; (ii) the definition of a set of questions to guide the recommendation process and obtain the information on relevant concerns from the user; (iii) evaluation of the elaborated set of questions used to drive the recommendation process; (iv) the Archify design and implementation, and (v) evaluation of the proposed system about the delivered result (the recommendation of architectural decisions). 

The result of Step (i) is presented in Section 2 of this paper, and the Archify design and implementation is in Section 3. Herein, we present the conduction of steps (ii) and (iii) in Section 4.1 and of Step (v) in Section 4.2.

For evaluating both the set of questions used in the recommendation process and the recommendations provided by the tool, we used survey research methodology inspired in the guidelines proposed by Lin\r{a}ker et al. \cite{Linaker:survey:guideline} and Seide et al. \cite{Molleri:2016:SGS:2961111.2962619}.
Therefore, the evaluation was performed following two survey researches: one for assessing \textit{(i) the relevance of the driving questions}; and (ii) another to evaluate the \textit{quality of the results returned by Archify}. Since our recommendation process and the survey are both formed by \textit{questions}, we distinct them herein as Recommendation Process Questions (RPQ) and Survey Questions (SQ). Both surveys were structured in four stages: (i) Planning, (ii) Execution, (iii) Analysis and (iv) Communication of results, as follows. 

\vspace{-0.2cm}
\subsection{Survey 1: The Relevance of the RPQ Set}

\noindent \textbf{Planning.} The recommendation process was composed of 12 RPQ. Those RPQ were designed to guide the process of choosing the most appropriate architectural alternatives and respective recommendation. A survey online form 
was elaborated with 20 SQ to be filled: one of agreement to participate the research, seven demographic SQ to collect information about the respondents, in addition to the 12 SQ designed to guide the process associated with 12 other text fields (one for each) so that respondents could make additional comments on each RPQ assessed. The Likert scale was adopted to allow respondents to judge the relevance of each question in regards to its relevance to assist in the process of recommending architectural solutions. Respondents could then choose the following values: Strongly disagree (1), disagree (2), neutral (3), agree (4), and strongly agree (5).
\\\\
\noindent \textbf{Execution.} The form was available between May 4th, 2020 and May 14th, 2020 (10 days). The invitation was sent to three groups of professionals by email and social media tools, reaching 484 people.
\\\\
\noindent \textbf{\underline{Results/Analysis.}} 19 responses were obtained, corresponding to a \textit{recall} of approximately 4\% responses. Next, an analysis of the collected data is performed. 

Regarding demographic aspects, 17 out of 19 respondents identify themselves as male and only two as female. Regarding the level of education, five respondents have not yet completed higher education (26.3\%), two have complete higher education (10.5\%), five professionals have MBA (26.3\%), six are masters (31.6\%) and one, a PhD (5.3\%). Among the respondents, 10 (52.6\%) are from industry and nine (47.4\%) were from academy. Among the 10 industry subjects, one had a double profile, working in both academia and industry.

About time experience on software architecture, development or software engineering, seven subjects (36.8\%) have zero to two years of experience, four people (21.1\%) have three to five years, two people from 6 to 9 years (10.5\%), and six of the respondents had (31.6\%) 10 or more years of experience. Hence, a significant part of the respondents had more than 10 years of experience.

Regarding the answers to this survey, for each of the RPQ elaborated, the respondents evaluated their relevance to assist in the architectural analysis and synthesis. Table \ref{tab:avaliacao1} shows the results for the evaluation of each RPQ. Each RPQ was assessed in regards to the respondents opinion if s/he Strongly Agrees (SA), Agrees (A), is Neutral (N), Disagrees (D) or Strongly Disagrees (SD) about the relevance of that question (and corresponding concern) to the architectural design process. Each column shows the percentage of responses obtained for each question and the last column shows the absolute percentage of people who answered Agree Fully or Agree (SA + A).

 We noticed that the respondents reacted positively to the questions asked. On average, the set of questions obtained 85.8\% of answers Strongly Agree or Agree, which reveals high relevance in the set of questions elaborated.  
  RPQ11 scored the lowest grade among the highest values. Some of them probably did not see a relation between usability and software architecture, although the literature reports results on the combination of both topics \cite{folmer2004architecting} and the relevance of usability within software architecture design. As it can be perceived, only the question RPQ7 caused a strong disagreement among the participants.  From the scores obtained by RPQ7 and RPQ11 obtained low evaluation, we perceived that the questions were not well formulated. We observed that the idea was not expressed properly and the respondent's understanding was not clear regarding the idea of these questions. We then reestructured the questions and could evaluate them again in the second survey, whose results are available in Section 4.2. The revised questions are presented in Table \ref{tab:perguntas}.
  We then reinforce a high acceptance of most of the elaborated RPQ.

\vspace{-0.6cm}
\begin{table}[!ht]
\label{tab:avaliacao1}
\caption{First survey results.}
\centering
\begin{tabular}{|l|l|l|l|l|l|l|}
\hline
\textbf{\#} & \textbf{SD} & \textbf{D} & \textbf{N} & \textbf{A} & \textbf{SA} & \textbf{A+SA} \\ \hline
RPQ1  & 0.0\%  & 5.3\%  & 0.0\%  & 47.4\% & 47.4\% & \textbf{94.7\%} \\ \hline
RPQ2  & 0.0\%  & 5.3\%  & 0.0\%  & 57.9\% & 36.8\% & \textbf{94.7\%} \\ \hline
RPQ3  & 0.0\%  & 0.0\%  & 10.5\% & 42.1\% & 47.4\% & \textbf{89.5\%} \\ \hline
RPQ4  & 0.0\%  & 5.3\%  & 5.3\%  & 42.1\% & 47.4\% & \textbf{89.5\%} \\ \hline
RPQ5  & 0.0\%  & 0.0\%  & 10.5\% & 73.7\% & 15.8\% & \textbf{89.5\%} \\ \hline
RPQ6  & 0.0\%  & 5.3\%  & 5.3\%  & 42.1\% & 47.4\% & \textbf{89.5\%} \\ \hline
RPQ7  & 10.5\% & 10.5\% & 26.3\% & 21.1\% & 31.6\% & \textbf{52.6\%} \\ \hline
RPQ8  & 0.0\%  & 0.0\%  & 0.0\%  & 57.9\% & 42.1\% & \textbf{100\%}  \\ \hline
RPQ9  & 0.0\%  & 5.3\%  & 10.5\% & 47.4\% & 36.8\% & \textbf{84.2\%} \\ \hline
RPQ10 & 0.0\%  & 0.0\%  & 5.3\%  & 52.6\% & 42.1\% & \textbf{94.7\%} \\ \hline
RPQ11 & 0.0\%  & 0.0\%  & 21.1\% & 47.4\% & 31.6\% & \textbf{78.9\%} \\ \hline
RPQ12 & 0.0\%  & 0.0\%  & 0.0\%  & 73.7\% & 26.3\% & \textbf{100\%}  \\ \hline
\end{tabular}
\end{table}
\vspace{-0.4cm}

\noindent \textbf{Final Considerations on the Evaluation 1.} The data collected and analyzed show a relevant acceptance by part of industry professionals and academics regarding the relevance of the questions designed to guide the recommendation process. 21\% of respondents \textit{strongly agreed} or \textit{agreed} with the relevance of all questions. Only four out of 19 disagreed with more than one question and three others disagreed with one question out of 13. 
In summary, we obtained an average of 85.8\% acceptance of the set of questions as a whole, which is positive. In the next section, we evaluate the recommendations delivered.

\subsection{Survey 2: The Archify Evaluation}

In the second evaluation, participants were invited to simulate a software development project and use the Archify tool to perform decisions about the raised concerns and analyse the results provided by the tool, i.e., the set of architectural decisions recommended.  
\\\\
\noindent \textbf{Planning.} 
A new evaluation form was made available (the second survey\footnote{Available here: https://archify.com.br/?survey2}). The form was composed of seven survey questions (SQ) for demographic identification, 12 SQ for evaluating the recommendations according to each question elaborated and other 12 text fields (one for each) so that the respondents could make additional observations, besides two questions to evaluate the result of the tool as a whole. The Likert scale was used again to allow respondents to judge the relevance of the recommendations.
\\\\
\noindent \textbf{Execution.} The online form as well as a link for respondents test Archify were available between July 21st, 2020 and July 24th, 2020 (4 days). The invitation was sent to a group of professionals by email and social media tools, totaling 89 people.
\\\\
\noindent \textbf{\underline{Reporting results.}} 19 responses were again obtained, approximately corresponding to a \textit{recall} of 28\%. Some of the respondents also participated in the same survey, and some of them were totally new to the tool. The results are available on an external link\footnote{Survey two results - https://bit.ly/3a0P7rv}.  17 out of 19 respondents are male and only two consider themselves female. About instruction level, six respondents have not yet completed higher education (31.6\%), two respondents have complete higher education (10.5\%), four have MBA (21.0\%), six masters (31.6\%) and one, PhD (5.2\%). 15 (79\%) came from the industry and four (21.0\%) from academia. Among the 15, four (21.0\%) had a double profile, working in both academia and industry. As for the time of experience with architecture, development or software engineering, three people (15.8\%) have up to two years of experience, four people (21.0\%) have three to five years, three people (15.8\%) from six to nine years and nine have 10 or more years of experience. Among the nine with 10 years or more of experience, three work in the industry and academia, five work exclusively in industry and one academic. 

\begin{figure}[!ht]
\centering
\includegraphics[width=0.6\textwidth]{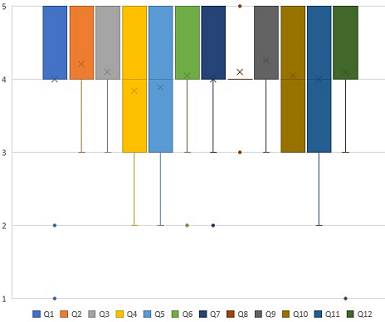}
\caption{Evaluation for each question recommendations.}
\label{fig:avaliacao-pergunta}
\end{figure}

The evaluation of the recommendations by question obtained an average score of 4.4 (scale from 1 to 5),  
revealing a high acceptance of the tool. It is worth mentioning that the three respondents with the major experience in the area (33, 20 and 18 years of experience) rated the software with grades 5, 4 and 5 respectively. As it can be perceived from Figure \ref{fig:avaliacao-pergunta}, the recommendations made by the tool received a good evaluation from the participants. In a Likert scale from 1 to 5 (Strongly disagree and Strongly agree with the relevance of the question and alignment between question and solution proposed), some respondents were neutral about questions RPQ4 (technology known by the team), RPQ5 (expertise on specific database), RPQ10 (Security) and RPQ11 (usability). Some outliers can be observed as disagreeing on questions RPQ6 (interoperability) and RPQ7 (Strong data typing). However, as it can also be observed in Figure \ref{fig:avaliacao-pergunta}, there is a predominance of strongly agree and agree with the relevance of the recommendations made by the system. These findings are complemented with qualitative analysis obtained and discussed, as follows.

For the purpose of assessing the relevance of the architectural decisions recommended by Archify, we also observed the comments let by the respondents in the survey form. Among the answers, we can highlight Respondent \#9, who has a master's degree and has worked at the academy for 10 years, stressed the possibility of using Archify's recommendations as documentation to support architectural decisions. He said: \textit{Respondent \#9: This guide is very interesting for a software architect to discuss with the entire team. I imagine that it also serves as documentation to support decisions that were made at the architectural level. }

Respondent \#11 has been in industry for 20 years and has a master's degree. S/he recommended the addition of a question to address software responsiveness and new possibilities for recommending technologies. S/he mentioned: \textit{Respondent \#11: I missed a question about responsiveness. I see that your proposal could evolve in the future to recommend other libraries and plugins, such as angular, bootstrap, jquery, ....}


Finally, Respondent \#13 mentioned the need for UI/UX improvement. In addition, he suggested making reference links available for greater knowledge of the decision in question. He said: \textit{Respondent \#13: I really liked the proposal, I don't know any other software that does this, so I found it very innovative and interesting. Suggestion of points to be improved: UI / UX, links with recommendations of articles or books in each section to better explain the suggested architecture and improvement of the images.}
\\\\
\noindent \textbf{Final Considerations on Evaluation 2.} From the data collected and analyzed, we observe a relevant approval by part of industry professionals and academics regarding the relevance of the tool. According to the assessment, 68.4\% of respondents rated Archify as excellent or good. In addition, 26.3\% considered themselves satisfied with the result presented. In other words, 94.7\% of the 19 participants approved the solution and the most experienced professionals were among the more satisfied.

\section{Discussion}

By triangulating results of two surveys, we can observe a high positive acceptance for both the set of questions elaborated and the architectural decisions recommended by Archify. 
Both academy and industry are largely aware of the importance of tools 
for bringing productivity and for for achieving quality in software development practice. Archify materializes knowledge on software architecture under a recommender system to small and medium-size teams that maybe do not have a software architecture expert among them.


 Limitations and threats to validity can certainly be discussed. Both survey-based evaluations were conducted with 19 professionals from academy and industry. This number of participants can be considered relatively low and the statistical force could be questioned. However, the group that participated in both studies were not identical, and the total number of different subjects could reach 30 different participants, which is a non-negligible number, which is also found in other survey studies of renowned conferences, as ICSE \cite{Sedano2019LTheProdBacklog} (27 participants). Hence, the threat to conclusion validity can be considered relieved. About internal validity (i.e., how the questions were selected to guide the recommendation process and how the subjects were selected), the decisions stored in the system were conceived from a set of two professionals with several years of experience with software architecture. To avoid the bias associated to the selection of the questions, the set of questions was assessed, revealing a high acceptance from participants. About the participants selection, they were randomly chosen, which relieves the threat related to bias. Moreover, the most experienced professionals evaluated the tool very positively, which reinforces its acceptance and relieves the threat. We did not proceed with hypothesis evaluation; hence, construction validity is not taken into account. For this scope, we do not generalize findings; hence, external validity is not considered.
 
 Regarding Archify limitations, we are aware it can still be further explored and expanded. Architectural decisions can (i) be conflicting, (ii) come from multiple sources, (iii) lead to a trade-off analysis and (iv) lead to multiple architectural alternatives \cite{paris2013difficulty}. In the current state, Archify does not cover a trade-off analysis among the involved QA yet, and do not recommend a specific combination between the triad architectural styles, technologies and architectural decisions recommended. Those characteristics can certainly be seen as future work initiatives, as the related work also do not cover such features. 
 
 As any other scientific endeavor, Archify also offers opportunities to further investigation and expansion. Nevertheless, Archify represents the vanguard of RS for SA. Besides being appreciated by experienced software architects with about 20-year experience, Archify gathers in a single tool all the most important functional characteristics of RS for SA found in correlated studies and tools. Archify offers an intuitive question-based process for users and has potential to assist software professionals in companies with recommendations on architectural decisions and technologies and exporting artifacts that can be used as part of the software product documentation. 
 In the next section, we present final remarks. 

\section{Final Remarks}
    The main contribution of this paper was to present the recommender system (RS) Archify as well as the results of its evaluation. Both the set of questions used to guide the recommendation process and the recommendations delivered were evaluated by professionals from academy and industry. Archify was emphatically supported by experienced practitioners and respondents with less experience were also very positive about it. Overall, we obtained an approval of 85.8\% on the set of questions used to guide the process and 94.7\% among 19 participants evaluating the tool results.
    
    From the feedback obtained, we enumerate future work, which include (i) the expansion of the tool knowledge base, (ii) definition and inclusion of other concerns (and corresponding questions), such as responsiveness, (iii) an empirical evaluation performed during a real software development project and (iv) adopting Archify in software architecture courses to assess how it could contribute to the acquisition and materialization of software architecture knowledge. 
    
    Regardless of the improvements suggestions, Archify advances the state of the art and practice by gathering several functionalities available in other related RS in a single tool. 
   It can support architects as a source of confirmation for their decisions founded on specialized literature, besides offering a comprehensive architectural documentation. We expect Archify can serve as a recommendation system to support small and medium-sized companies in the pursue for software quality assurance.
  
\bibliographystyle{splncs04}  
\bibliography{bibliography}
\end{document}